\documentstyle[psfig]{cupconf}

\ifoldfss
\else
  \ifnfssone
    \newmathalphabet{\mathit}
      \addtoversion{normal}{\mathit}{cmr}{m}{it}
      \addtoversion{bold}{\mathit}{cmr}{bx}{it}
    \newmathalphabet{\mathcal}
      \addtoversion{normal}{\mathcal}{cmsy}{m}{n}
    \else
    \ifnfsstwo
    \fi
  \fi
\fi

\def\hexnumber#1{\ifcase#1 0\or1\or2\or3\or4\or5\or6\or7\or8\or9\or
 A\or B\or C\or D\or E\or F\fi }

\makeatletter
\ifx\CUP@mtlplain@loaded\undefined
\else
\fi
\makeatother
 \makeatletter
 \ifx\CUP@mtlplain@loaded\undefined
   \font\tenbmi=cmmib10 at 10pt
   \font\sevenbmi=cmmib10 at 7pt
   \font\fivebmi=cmmib10 at 5pt

   \newfam\bmifam
   \textfont\bmifam=\tenbmi
   \scriptfont\bmifam=\sevenbmi
   \scriptscriptfont\bmifam=\fivebmi
   
 \fi
 \makeatother
\ifnfsstwo

\fi
\ifnfssone

\fi
\ifoldfss

\fi

\mathchardef\varLambda="0103

\makeatletter
\ifx\CUP@mtlplain@loaded\undefined
\else
\fi
\makeatother
\makeatletter
\ifx\CUP@mtlplain@loaded\undefined
  \font\tenbms=cmbsy10
  \font\sevenbms=cmbsy10 at 7pt
  \font\fivebms=cmbsy10 at 5pt
  \newfam\bmsfam
  \textfont\bmsfam=\tenbms
  \scriptfont\bmsfam=\sevenbms
  \scriptscriptfont\bmsfam=\fivebms

  \edef\bsy@{\hexnumber\bmsfam}
  \mathchardef\bnabla="0\bsy@72
\fi
\makeatother
\title[Slow Jet in NGC 1068]{Slow Jets in Seyfert Galaxies: NGC 1068}

\author[A. L. Roy {\it et al.\/}]%
{
A.\ns L.\ns R\ls O\ls Y$^1$
A.\ns S.\ns W\ls I\ls L\ls S\ls O\ls N$^{2,3}$
J.\ns S.\ns U\ls L\ls V\ls E\ls S\ls T\ls A\ls D$^4$
E.\ns J.\ns M.\ns C\ls O\ls L\ls B\ls E\ls R\ls T$^5$
}

\affiliation{$^1$Max-Planck-Institut f\"ur Radioastronomie, Auf dem H\"ugel 69, 
D-53121 Bonn, Germany\\[\affilskip]
$^2$ Department of Astronomy, University of Maryland, College Park, MD
20742, USA\\[\affilskip]
$^3$ Space Telescope Science Institute, 3700 San Martin Drive, Baltimore, 
  MD 21218, USA\\[\affilskip]
$^4$ National Radio Astronomy Observatory, P.O. Box O, Socorro, NM
87801, USA\\[\affilskip]
$^5$ Johns Hopkins University, Department of Physics and Astronomy, Homewood Campus, 3400 North Charles Street, Baltimore, MD  21218, USA\\[\affilskip]
}

\begin{document}
\ifnfssone
\else
  \ifnfsstwo
  \else
    \ifoldfss
      \let\mathcal\cal
      \let\mathrm\rm
      \let\mathsf\sf
    \fi
  \fi
\fi

\maketitle

\begin{abstract}

We have used the Very Long Baseline Array at 5 GHz to image the
nucleus of NGC 1068 at two epochs separated by 2.92 yr.  No relative
motion was detected between the high brightness-temperature knots
within components NE and C relative to the nuclear component S1,
placing an upper limit of 0.075 $c$ on the relative component speeds
at distances of 21 pc and 43 pc from the AGN.  The low speed is
consistent with the low bulk flow speed previously inferred from
indirect arguments based on ram pressure at the bow shock and on line
emission from the jet-cloud collision at cloud C.  The components 
are probably shocks in the jet, and the bulk flow speed could conceivably
be higher than the limit reported here.

\end{abstract}

\firstsection 
\section{Introduction}

Four years ago we began our VLBA studies of nearby Seyfert galaxies
with the aims of imaging the central radio sources, looking for
eabsorbed core spectra that might be due to an obscuring torus (Roy {\it et
al.} 1999), looking for direct thermal emission from ionized gas at
the inner edge of the torus (Mundell {\it et al.} 2000), and measuring jet
speeds for comparison to the jets in powerful radio galaxies and
quasars (Ulvestad {\it et al.} 1999).  Enough time has now elapsed to detect
motions slower than 0.1 $c$ (depending on distance) and we have made
relative speed measurements on scales of parsecs to tens of parsec
scales in four Seyferts (Mrk 231, Mrk 348, NGC 4151, and NGC 5506).
These have turned out to be systematically slow ($< 0.25$ $c$)
compared to jets in powerful radio galaxies and quasars, 
which is interesting when
one attempts to distinguish between intrinsic and extrinsic effects
that cause jets in Seyfert galaxies to be weak.  In this
paper we report an upper limit on the motion of compact components in
NGC 1068.

\section{NGC 1068 Background}

NGC 1068 is a nearby (14.4 Mpc) archetypical type 2 Seyfert with a
hidden broad-line region.  Within its 13 kpc radio disc emission is
a 1 kpc-long central linear structure, which radiates $3.6 \times
10^{22}$ W Hz$^{-1}$ at 4.9 GHz.  VLA observations at 15 GHz 
with 0.4 arcsec resolution (Wilson \& Ulvestad 1987) resolve
the linear structure into a classic core-jet-lobe structure, and from
ram-pressure arguments at the bow shock, they infer that the
incident jet is slow (4000 to 15000 km s$^{-1}$) and 
light (2.5 $\times$ 10$^{3}$ to 10$^{5}$ m$^{-3}$).
MERLIN observations at 5 GHz
with 60 mas resolution (Muxlow et al. 1996) resolve considerable
detail in the jet, and spectral ageing rates along the jet infer a jet
velocity $< 5 \times 10^4$ km s$^{-1}$ (Gallimore et al. 1996b).  A
water-maser disc is present around component S1 (Gallimore et
al. 1996a; Greenhill et al. 1997) and the 
rotation curve infers a mass within the central 0.65 pc
of $1.5 \times 10^7 M_\odot$.  The accretion rate is
$0.04$ $M_\odot$ yr$^{-1}$ from the maser properties (Maloney et
al. 1997, private communication to Bicknell et al. 1998) and 0.05
$M_\odot$ yr$^{-1}$ from the bolometric luminosity, assuming a 
conversion efficiency of accreted mass into bolometric luminosity of 0.1 
(Bicknell et al. 1998).

A detailed analysis of the jet-cloud collision at component C was
carried out by Bicknell et al. (1998).  Assuming that the optical line
emission from the cloud is excited by a radiative shock driven by the
jet impact, they found the mass flux incident on the cloud is 0.5
$M_\odot$ yr$^{-1}$, which is $10\times$ larger than the accretion
rate, and that the incident jet speed is 0.04 $c$ (within
a factor of three).  They found such a slow,
mostly thermal jet is different from the ultra-relativistic
plasmas found
in powerful radio jets, and suggest that Seyfert jets may be
radio-weak because they carry a substantial load of thermal plasma,
perhaps due to entrainment or due to the jet formation mechanism.  It
is therefore of considerable interest to make a direct measurement of
the jet speed in NGC 1068.

\section{Observations}

NGC 1068 was observed using the VLBA at 5.0 GHz at two epochs
separated by 2.92~yr between 1 June 1996 and 3 May 1999.  We recorded
LCP with two-bit sampling and 32 MHz bandwidth
during the first epoch, and the same but 64~MHz bandwidth during the
second epoch.  The data were initially phase calibrated by phase
referencing to the nearby source J0239-0234, which provided a
sufficiently good starting model for phase self-calibration on NGC
1068.  Final images were made using uniform weighting.  Baselines
longer than 40 M$\lambda$ were discarded because those did not detect
NGC~1068, yielding beam sizes of 6.2~mas $\times$ 4.3~mas for the
first epoch and 5.9~mas $\times$ 4.4~mas for the second epoch.  Both
epochs were processed in the same manner at the same time by the same
person (ALR), using AIPS.  On-source integration times were 1.2~hr in
the first epoch and 2.4 hr in the second epoch, yielding rms noise of
0.20 mJy beam$^{-1}$ for the first-epoch image and 0.16 mJy
beam$^{-1}$ for the second-epoch image, which are close to the thermal
noise limit.  Positions were measured in the image plane by fitting a
quadratic surface to each peak.  The formal uncertainty of the peak
position was estimated as equal to the synthesised beamwidth divided
by the signal-to-noise ratio for each component, and the individual
uncertainties were combined in quadrature to estimate the uncertainty
in component separation.  For example, for components S1 and C in the
first epoch, this yielded a 1 $\sigma$ uncertainty of 0.32 mas for the
component separation.

\section{Results}

We detected at least three compact, bright components (Fig 1) that lie
within the extended regions S1, C and NE of the MERLIN image by Muxlow
et al. (1996).  Most of the extended jet emission seen in the MERLIN
image is not visible in the VLBA images because of their much smaller
beam and hence poorer brightness sensitivity limit.

\begin{figure} 
  \vspace{0pc}
  \psfig{file=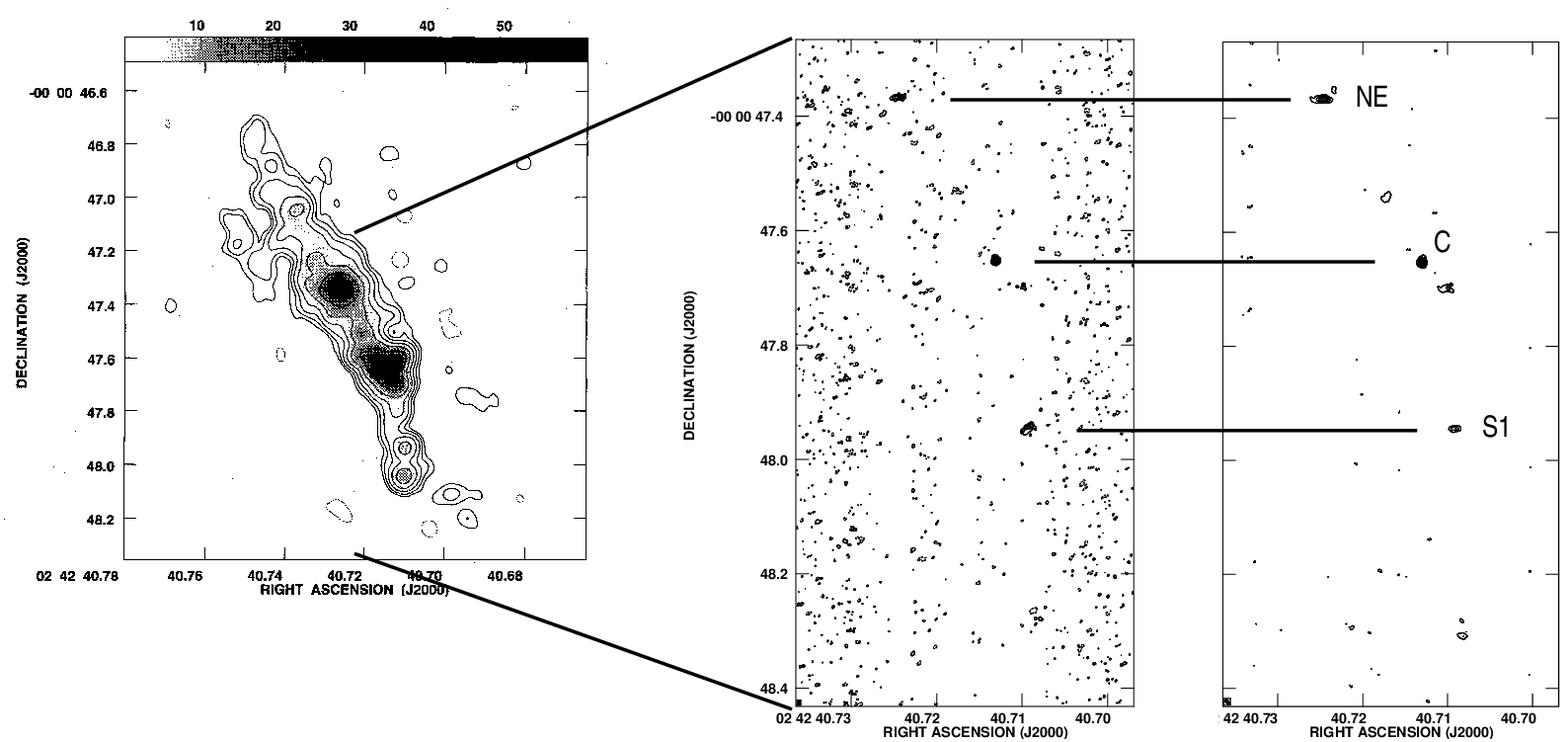,width=13.5cm}
  \caption{Left: MERLIN 5.0 GHz image by Muxlow et al. (1996) for
  orientation, and the VLBA images of NGC 1068 at 5.0 GHz at epochs
  1996.42 (middle) and 1999.34 (right).  Contour levels are -1.5, -1.0,
  -0.5, 0.5, 1.0, 1.5, ... 5.5 mJy beam$^{-1}$ and the restoring beam is
  6.2 mas $\times$ 4.3 mas in position angle $-17^\circ$ in 1996.42 and 
  5.9 mas $\times$ 4.4 mas at $48^\circ$ in 1999.34}
\end{figure} 

The relative component separations (Table 1) did not change significantly
between the two epochs.  The maximum difference in component separation 
was 0.3 mas, which is 0.07 of a beamwidth and is too small to be significant.
We quote a 3 $\sigma$ upper limit to the separation change of 0.96 mas which
corresponds to $v < 0.075$ $c$ (assuming a distance to NGC 1068 of 14.4 Mpc 
for $H_{0}$ = 75 km s$^{-1}$ Mpc$^{-1}$).

\begin{table} 
  \begin{center} 
  \begin{tabular}{ccccccc} 

      {\it Epoch} & {\it Component} & {\it dRA} & {\it ddec} & {\it total} & {\it peak flux density} & {\it total flux density} \\[3pt]
                  &                 &   (mas)   &    (mas)   &    (mas)    &   (
mJy beam$^{-1}$)     &     (mJy)\\

\\
       1996.42 &  NE &   230.7  & 577.3 &  621.7  &   4.2    &         $11.4 \pm 2.3$ \\
               &  C  &    59.0  & 289.5 &  295.5  &  12.1    &         $24.2 \pm 4.3$ \\
               &  S1 &     0.0  &   0.0 &    0.0  &   3.3    &         $ 4.5 \pm 1.2$ \\

\\

       1999.34 &  NE &   228.3  & 577.9 &  621.4  &   4.3    &         $ 8.2 \pm 1.6$ \\
               &  C  &    59.4  & 289.8 &  295.8  &   9.2    &         $17.3 \pm 3.0$ \\
               &  S1 &     0.0  &   0.0 &    0.0  &   2.4    &         $ 5.5 \pm 1.2$ \\

  \end{tabular}
  \caption{Preliminary measurements of NGC 1068 component parameters from Fig 1}
  \end{center} 
\end{table}

\section{Discussion}

The upper limit to the component speeds in NGC 1068, $v < 0.075$ $c$,
is consistent with the low jet speeds
inferred by both Wilson \& Ulvestad (1987) (0.013 $c < v <$
0.050 $c$) and by Bicknell {\it et al.} (1998) (0.04 $c$ with an
uncertainty of a factor of three).  Note however that we may have
measured the speed of shocks within the flow, and the flow itself
might be faster.  Measurement of the large-scale bulk flow must await
a proper-motion measurement of the extended emission visible to
MERLIN.  However, to detect motion of $0.1\times$ the MERLIN 60 mas
beam at 5 GHz (0.42 pc) with a flow speed of, say, 0.05 $c$, would
require 27 years between epochs, or more due to limited image fidelity.

Proper-motion measurements or limits are available now for six Seyfert
galaxies, and the distribution of speeds is shown in Fig 2 along with
a histogram of speeds in powerful radio sources for comparison.  The
Seyferts clearly have relatively slow jets on the parsec scale, and
thus we conclude that Seyfert galaxy jets must be either launched
sub-relativistically, or else are decelerated within the first parsec
or ten parsecs.

\begin{figure} 
  \vspace{0pc}
  \psfig{file=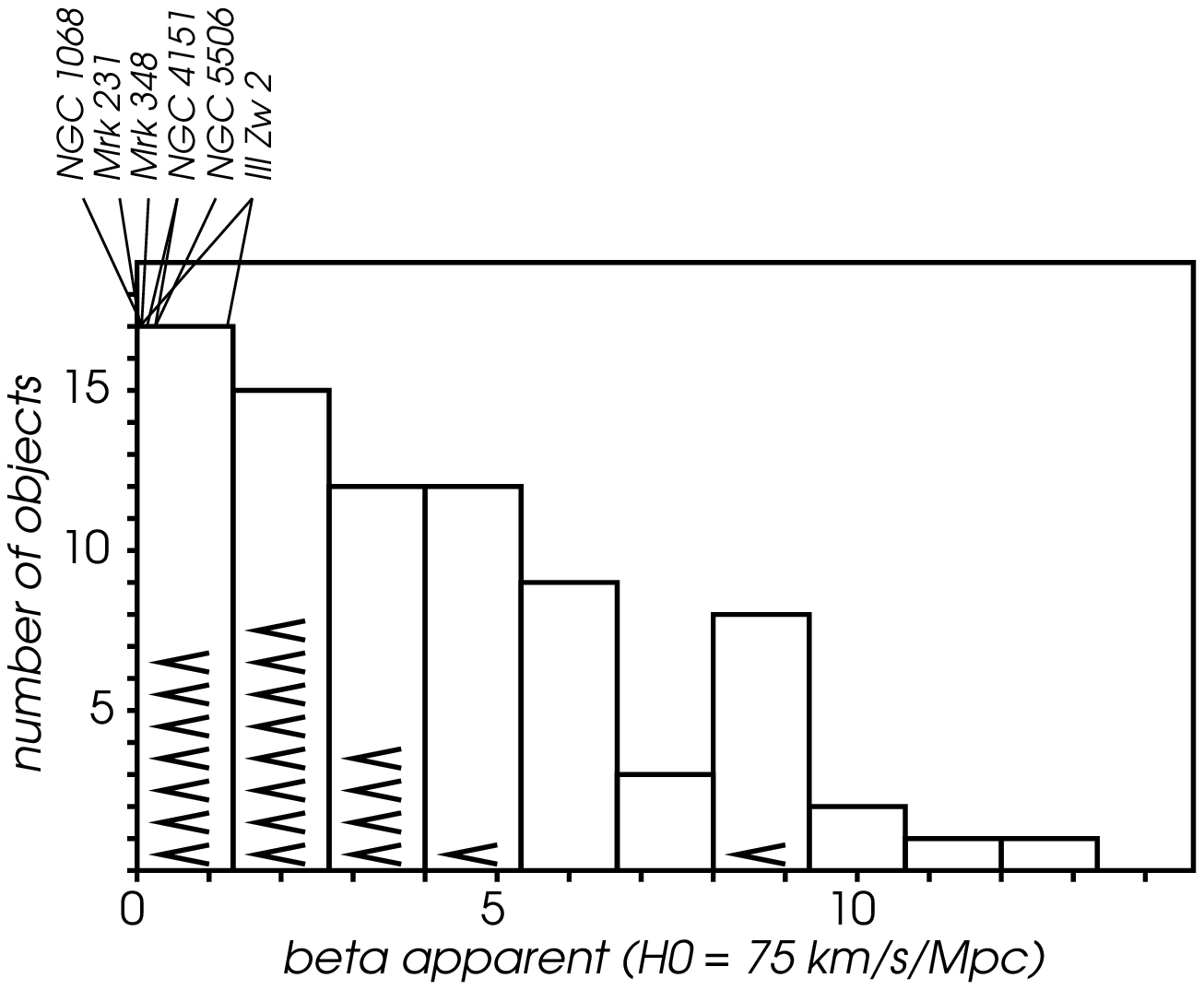,width=10cm}
  \caption{The distribution of jet speeds of Seyfert galaxies with
  proper motion or upper limits measured (III Zw 2 is from
  Brunthaler {\it et al.} 2000) and, for comparison, is shown a uniform
  sample of strong, flat-spectrum objects, comprised of quasars, 
  BL~Lac objects, galaxies, and empty fields (Vermeulen 1995).  The
  Seyferts populate the slow end of the distribution.}
\end{figure}

\section{Conclusion}

We have made VLBI images of the nucleus of NGC 1068 at 5 mas
resolution and at two epochs spaced by 2.92 yr, and have detected no
motion of the compact components greater than 0.075 $c$ (3 $\sigma$),
consistent with speeds inferred by Wilson \& Ulvestad (1987) and
Bicknell et al. (1998).  The compact components may be standing shocks
in the jet and the large-scale flow could conceivably be faster.
Sub-relativistic jet speeds have been found now in six Seyfert
galaxies.  Future measurements can lower the upper limits to the
speeds of the compact components.  However, measurements with lower
resolution to measure motion of the large-scale outflow would require
about a 30 yr baseline to provide interesting limits on NGC 1068.
Better targets might lie at higher declinations where better image
fidelity is achievable.

\begin{acknowledgments}
ALR enjoyed discussions with A. Pedlar, T. Muxlow, and C. Mundell 
in the charming English countryside, about Seyferts and
proper motion measurements with MERLIN.
\end{acknowledgments}

\end{document}